\def\bq{\begin{equation}}
\def\eq{\end{equation}}
\def\bqa{\begin{eqnarray}}
\def\eqa{\end{eqnarray}}
\def\bqb{\begin{eqnarray*}}
\def\eqb{\end{eqnarray*}}
\documentstyle[12pt]{article}\textwidth 16cm
\def\pr#1#2#3{ Phys. Rev. ${\bf{#1}}$ (#2) #3 }

\def\pl#1#2#3{ Phys. Lett. ${\bf{#1}}$ (#2) #3 }

\def\np#1#2#3{ Nucl. Phys. ${\bf{#1}}$ (#2) #3 }
\def\zp#1#2#3{ Z. Phys. ${\bf{#1}}$ (#2) #3 }

\def\ie{{\it i.e.\/}}

\def\L{ {\cal L }}
\def\O{ {\cal O }}

\def\swd{s^2_W}
\def\cwd{c^2_W}
\def\cot{\cos\theta}
\def\sit{\sin\theta}
\def\ed{e^2}
\def\mwd{M^2_W}
\def\lw{\lambda_W}
\def\rd{\sqrt2}
\def\mh2{m^2_H}
\def\mzd{M^2_Z}

\def\roughly#1{\mathrel{\raise.3ex
    \hbox{$#1$\kern-.75em\lower1ex\hbox{$\sim$}}}}
\def\lsim{\roughly<}

\begin{document}
\pagenumbering{arabic}
\thispagestyle{empty}

\begin{flushright} PM/94-13 \\ THES-TP 94/03 \\ April 1994
 \end{flushright}
\vspace{1cm}
%---------------------titre ---------------------------------------
\begin{center}
{\Large\bf Higgs Production}\\
{\Large\bf in $SU(2)_c$ Symmetric Interactions}
\footnote{Work supported by the scientific cooperation program
 between CNRS and EIE.}

 \vspace{2.0cm}
%-----------------------------------------------------------------
 {\Large  G.J. Gounaris } \\
 Department of Theoritical Physics \\
University of Thessaloniki, Greece
\vspace {0.5cm}  \\
 {\Large  J. Layssac, F.M. Renard }\\
 Laboratoire de Physique
Math\'{e}matique \footnote {Unit\'{e} Associ\'{e}e au CNRS $n^0$
040768.}\\
Universit\'{e} de Montpellier II Sciences et
Techniques du Languedoc \\
Place E.Bataillon, Case 50 F-34095 Montpellier Cedex 5,
France  \\
\vspace {2cm}

 {\bf Abstract}
\end{center}
\noindent
We study the sensitivity of Higgs production and decay processes
to the $SU(2)_c$ symmetric couplings $O_W$ and $O_{UW}$.
Remarkable results are obtained in the
case of
$\gamma \gamma \to H$ and for certain ratios of Higgs decay widths.
We also discuss and complete previous results on
unitarity constraints
for such couplings.

\vspace{1cm}

\setcounter{page}{0}

\clearpage

\section{Introduction}

Future $e^+e^-$ and pp colliders \cite{nlc, lhc, phys}
offer many possibilities for
testing the bosonic sector of
the electroweak interactions through vector boson pair production.
Such possibilities have been studied in \cite{VVVV,uni,LHCGLR,nlcgr},
where guided by the principle of $SU(2)_c$ symmetry \cite{SU2c},
supported by present
tests at LEP1 and at low energy experiments, we considered
the New Physics (NP) effects generated by the
operator\\
\bq \O_W={1\over3!}\left( \overrightarrow{W}^{\ \ \nu}_\mu\times
  \overrightarrow{W}^{\ \ \lambda}_\nu \right) \cdot
  \overrightarrow{W}^{\ \ \mu}_\lambda =-{2i\over3}
\langle W^{\nu\lambda}W_{\lambda\mu}W^\mu_{\ \ \nu}\rangle \ \ \
, \eq\\
with a coupling $\lambda_W$,
 and the operator\\
\bq \O_{UW}=\langle (UU^{\dagger}-1)\ W^{\mu\nu} \
W_{\mu\nu}\rangle \ \ \ , \ \ \eq\\
with
a coupling called d. In these operators the definitions\\
\bq U=\bigm(\widetilde \Phi\ \ , \ \Phi\bigm){\sqrt2\over v} \ \ \ \
 \ \ \eq\\
are used,
where  $\Phi$ is the standard Higgs
doublet, $v$ its vacuum expectation value,   $\widetilde
\Phi = i\tau_2 \Phi^*$,
and  $\langle A \rangle \equiv TrA$. Thus,
the first of these operators involves gauge
bosons only, whereas
the second one includes Higgs bosons also. We have found that
the coupling $\lambda_W$ affects $e^+e^-$ and $q\bar q$ annihilation
processes to gauge boson pairs, as well as boson-boson fusion
through 3-boson
and 4-boson vertices.  From the analysis of
gauge boson pair production it was concluded \cite{Bil,VVVV}
that a sensitivity
on $|\lambda_W|$ of 0.002 (0.01) could be reached
at $e^+e^-$ (pp) colliders. Correspondingly for the d coupling,
we have found that the $\O_{UW}$ contribution to the gauge boson pair
production arises only
through Higgs
exchange diagrams in boson fusion processes, thereby
leading to much weaker sensitivities on $|d|$, \ie\,
0.02 (0.1) from $e^+e^-$ (pp), \cite{nlcgr,VVVV}.\par

The first purpose of this paper is to show that the $\O_{UW}$
sensitivity can be improved by looking also at Higgs
production and decay
processes. Assuming that the Higgs particle
is sufficiently light to be  actually produced,
one expects a
particularly high sensitivity in those processes
where the standard
contribution is suppressed. Such processes are those determined
by the $H\gamma\gamma $ and  $H\gamma Z$ couplings,
which receive standard contributions only at
1-loop, whereas their contributions from $\O_{UW}$
already arise at tree level.
Since one would expect large $d$ effects in amplitudes
involving these couplings, we
pay a special attention to Higgs production via $\gamma\gamma$
collisions. In addition we also consider the other main H production
mechanisms, namely through gauge boson fusion (\ie , WW, ZZ,
 $\gamma \gamma$ and $\gamma Z$ collisions)
and through associate
production in $e^+e^- \to HZ$ and $q \bar q \to HW$ at pp colliders.
Finally, the d sensitivity achievable
by measuring the Higgs decay modes  $H \to \gamma \gamma$
and $H \to \gamma Z$, is also studied.\par

The second aim of the present work is to compare the
aforementioned sensitivities,
to more theoretical constraints on $\lambda_W$ and $d$.
One indirect way of deriving such constraints, is by using LEP1
measurements in 1-loop considerations
involving Higgs exchange diagrams \cite{ZepHiggs}.
In a different approach, very stringent constraints may be derived
in a purely theoretical way, by using unitarity \cite{uni}.
These arise because of the high dimensionality of the operators
$\O_W$ and $\O_{UW}$,
which necessarily leads to violation of unitarity above a certain scale.
Thus, an assumption on the magnitude of the scale below which no
strong interactions appear (\ie\, no unitarity saturation), immediately
implies an upper bound on $|\lambda_W|$ and $|d|$.
These limits were established in \cite{uni} using the
gauge boson-gauge boson scattering amplitudes given in
\cite{LHCGLR}, and the Higgs involving amplitudes  presented
here. The later  are
required for establishing the unitarity bounds on $d$, and
they can also be useful for estimations of the Higgs production
rates. Morever, here we confirm the results of the previous letter
\cite{uni}, where the unitarity constraints on $d$ were derived
on the basis of the $J=0$
partial waves amplitudes only, by looking also at the $J=1$
partial waves.\par

The contents of the paper is the following. Sect.2 is devoted to
Higgs production from photon-photon collisions using laser
backscattering in a high energy $e^+e^-$ collider. In Sect.3 we
consider Higgs production through  gauge boson
fusion processes at $e^+e^-$ colliders, using the
Weisz\"acker-Williams approximation, and in Sect.4
the associate
production processes  $e^+e^- \to HZ$ and $e^+e^- \to H
\gamma$ are studied. The $d$ sensitivity to the Higgs decay modes
is studied in Sect.5, and the unitarity constraints in Sect.6.
Finally the last Sect.7
resumes the experimental and theoretical prospects about the
$d\O_{UW}$ coupling. Appendix A and B give the explicit
expressions for the $\lambda_W \O_W$ and $d\O_{UW}$
contributions to the single and double Higgs production
amplitudes at an energy above 1TeV.\par

\section{Higgs production in photon-photon collisions from laser
backscattering}

The Standard Model (SM) contribution to the $H\gamma\gamma$ coupling is
rather weak as it only
appears through 1-loop \cite{hgg}. Nevertheless owing
to the large $\gamma $ luminosities that may be available
for double laser
backscattering on the high energy $e^{\pm}$ beams, a copious Higgs
production would be expected in $\gamma \gamma $ collisions
\cite{nlc, phys}. Such a production must be very senstive to
the $d\O_{UW}$ interaction, which contributes to it at tree
level.\par

The predominant SM contributions to the $H\to\gamma \gamma$
width, are due to the top and W loops \cite{hgg}. Adding to this the
tree level
$d\O_{UW} $ contribution, we obtain

\bq   \Gamma(H\to\gamma \gamma) = {\sqrt2 G_F \over 16\pi} m^3_H
\left [{\alpha\over4\pi}({4\over3}F_t + F_W) - 2ds^2_W
\right ]^2 \ \ \ \ \ \ , \ \ \ \ \ \eq \\
where the top and $W$ contributions are respectively determined
by\\
\bq   F_t = -2t_t(1+(1-t_t)f(t_t))   \ \ \ \ \ , \ \ \ \ \ \eq

\bq   F_W = 2+3t_W+3t_W(2-t_W)f(t_W)) \ \ \ \ \ \ , \ \ \ \ \eq \\
in terms of  \\
\bqa
f(t) = \left [sin^{-1}(1/\sqrt{t})\right]^2 \ \ \ \ \ \ \ \
\ \ &
\makebox{\ \ \ \ \ if \ \ \ } &
           t \geq 1 \ \ \ \ , \ \ \ \nonumber \\[0.5cm]
f(t)= -{1\over4}\left [\ln \left ({1+\sqrt{1-t} \over 1-\sqrt{1-t}}
\right )
-i\pi\right ]^2 & \makebox{\ \ \ \ if \ \ \ } & t < 1 \ \
\ \ , \ \ \ \ \eqa
where $t_t =4m^2_t/\mh2$\, , \, $t_W=4\mwd/\mh2$\, .
Using the width $\Gamma(H \to \gamma \gamma)$,
the cross section
 $\sigma^H_{\gamma \gamma}$   for the
elementary process $\gamma \gamma \to H$ is then expressed as\\
\bq    \sigma^H_{\gamma \gamma} = {8\pi^2 \over
 m_H}~\Gamma(H\to\gamma \gamma) \delta(s_{\gamma \gamma}-\mh2)
\ \ \ \ \ \ .  \ \ \ \ \ \ \ \eq \par

In order to calculate now the corresponding
cross section for double laser scattering in  $e^+e^-$ colliders,
we need the induced spectral $\gamma
\gamma$ luminocity. For unpolarized laser and $e^{\pm}$
beams, this is given by \cite{laser}\\
\bq  \frac{d\L_{\gamma \gamma}(\tau)} {d\tau} ~ =~  \int ^{x_{max}}
_{\tau\over{x_{max}}}\ {dx\over{x}}\, f^{laser}_{\gamma/e}(x)\,
f^{laser}_
{{\gamma /e}}\left({\tau \over{x}}\right) \ \ \ \ \ \ \ \ \
\ \ \ , \ \ \ \   \eq\\
where
\bq \tau = {s_{\gamma \gamma} \over s_{ee}} \ \ \ \ \ \ \ \ \ \ \
 \ \ \ \ \ \ \    \eq\\
\noindent is the ratio
of the $\gamma \gamma $ c.m. squared energy to the $e^+e^-$ one,
 $\xi=2(1+ \sqrt{2}) \simeq 4.8$ and
 $x_{max}=\xi/(1+\xi)\simeq 0.82$ \cite{fgamma,Kuhn}.
The photon distribution $f^{laser}_{{\gamma /e}}(x)$ in (6) is
obtained by Compton scattering the laser beam on the $e^\pm$
beams and it is given by \cite{fgamma, Kuhn}\\
\bq
f^{laser}_{{\gamma /e}}(x)~ =~{1\over{D(\xi)}}
\left(1-x+{1\over{1-x}}-{4x\over{\xi(1-x)}}
+{4x^2\over{\xi^2(1-x)^2}}\right)
  \ \ \ ,\ \   \eq
where $x$ is the fraction of the incident $e^{\pm}$ energy carried by
the backscattered photon, while \\
\bq      D(\xi) =
\left(1-{4\over{\xi}}-{8\over{\xi^2}} \right) \ln (1+\xi)\, +\,
{1\over2}+{8\over{\xi}}\,-\,
{1\over{2(1+\xi)^2}}
  \ \ \ .\ \
\eq\\ \par

Using the above $\gamma \gamma$ luminosity, the cross
section for Higgs production through double
laser scattering is then given by \\
\bq
{d\sigma\over d\tau}~ = ~{d\L_{\gamma
\gamma}(\tau)\over d \tau}~
 \sigma^H_{\gamma \gamma}   \ \ \ \ \  , \ \ \ \
\eq\\
which combined with (8) leads to the integrated cross section\\
\bq   \sigma  = \L_{\gamma \gamma}(\tau_H)~\left({8\pi^2 \over
 m_H} \right)~{\Gamma(H\to\gamma \gamma)\over s_{ee}}
    \ \ \ \ \ \ \ . \ \ \ \ \ \ \eq
The correspondingly expected number of events per year
is then determined by\\
\bq N~ =~ \bar {\cal L}_{ee}\ \sigma \ \ \ \ \ \ , \ \ \ \ \ \ \
   \eq
in terms  of the integrated $e^+e^-$ annual
luminosity ${\bar\L}_{ee}$ taken to be $10 fb^{-1}
year^{-1}$.\par

The results are presented at variable Higgs mass
in Fig.1a-c for a 0.5, 1 and 2 TeV center of mass
energy $e^+e^-$ collider.
 One recognizes the typical structure of the standard predictions with
the dips and bumps due to the $t \bar t$ and $W^+W^-$ threshold
effects in the loop and the
interferences of their contributions. As expected the
sensitivity to $d$ is very
interesting. With the predicted number of events,
values of $d$ down to $\pm 0.001$ could even be observed.
As one can be see on
the figures, the precise
value of the
observability limit strongly depends on the
Higgs mass, which determines the important interferences between
the SM and $\O_{UW}$ contributions.\par

\section{Higgs production by WW or ZZ fusion}

 Another standard way of producing the Higgs boson in $e^+e^-$ or pp
colliders is through fusion of the vector bosons emitted from
the fermions. Let us first discuss the
processes $e^+e^- \to \nu \bar\nu (WW) \to \nu \bar\nu H$ and
$e^+e^- \to e^+e^- (V_1V_2) \to e^+e^- H$.
In SM, the first one goes through $W^+W^- \to H$, while
at tree level
the second one goes through $ZZ \to H$.
The corresponding SM cross sections
have been given in a compact form for $V=W$ or $Z$ in \cite{abdel}.
In order to include the  $d\O_{UW}$ effect, one has
to separate the
transverse-transverse (TT) from the longitudinal-longitudinal
(LL) WW
states. Because of the gauge invariant nature of the $\O_{UW}$
operator, only the TT partial width is appreciably affected by it,
as one can see
from the subsequent expressions for the Higgs partial widths\\
\bq \Gamma(W_TW_T)~=~ \left( {\alpha \mwd \beta_W
\over 2\swd m_H}\right)~
\left [1-{d \over \mwd}(\mh2-2\mwd)\right]^2 \ \ \ \ \ , \ \ \
\ \ \ \ \     \eq\\
\bq  \Gamma(W_LW_L)~=~ \left ({\alpha \beta_W
\over 4 \swd m_H} \right )~
\left[{(\mh2-2\mwd)^2 \over 4\mwd}+4d^2\mwd-2d(\mh2-2\mwd)\right ]\ \
\ \ \ , \ \ \
\ \   \eq\\
where $\beta_W=\sqrt{1-4\mwd /\mh2}$ is the velocity of the final
state W's. We note that the total $H \to W^+W^-$ width \\
\bqa \Gamma_{tot}(WW) = \ \ \ \ \ \ \ \ \ \ \ \ \ \ \ \ \ \ \ \
\ \ \ \ \ \ \ \ \ \ \ \ \ \ \ \ \ \ \ \ \ \ \ \ \ \ \ \ \ \ \ \
\ \ \ \ \ \ \ \ \ \ \  \ \
\ \ \ \ \ \ \ \ \ \ \ \ \ \ \ \ \ \nonumber\\[0.2cm]
  \left( {\alpha  \beta_W \over 4 \swd m_H}\right )
\left [2\mwd+ {(\mh2-2\mwd)^2\over 4\mwd}
 + 4d^2\left \{{(\mh2-2\mwd)^2\over 2\mwd}+\mwd \right\}-
6d(\mh2-2\mwd) \right ]
\ ,  \eqa\\
being dominated by the LL state, is relatively less affected
by d, than the TT part given in (16).\par

The $e^+e^- \to e^+e^- (V_1V_2) \to e^+e^- H$ case
is more delicate. As mentioned above in SM it is dominated by
$ZZ \to H $, but when  $d\O_{UW}$ effects are
included, contributions from  the $\gamma \gamma \to H$ and
$\gamma Z \to H$ fusion processes
should be added. In the spirit of the
Weisz\"acker-Williams approximation used here, we need therefore
the corresponding Higgs partial widths to these three channels.
For the later two, we keep
and 1-loop SM contributions, together with tree level
$d\O_{UW}$ effects.\par

The width for $H \to \gamma \gamma$ is already
given in (4). The corresponding expession for $H \to \gamma Z$,
based on the
standard 1-loop (top and W) amplitudes \cite{hgg} and the
tree level $\O_{UW}$
contribution, is\\
\bq   \Gamma(H\to \gamma Z) = {\sqrt2G_Fm^3_H\over 32\pi}(1-{\mzd
\over \mh2})^3|{\alpha\over2\pi}(A_t+A_W)+4ds_Wc_W|^2 \ \ \ , \ \
\eq\\
\bq   A_t =
{(-6+16s^2_W)\over 3s_Wc_W}[I_1(t_f,l_t)-I_2(t_t,l_t)]
\ \ \ \  , \ \ \ \ \ \ \eq\\
\bq
A_W=-cot\theta_W[4(3-tan^2\theta_W)I_2(t_W,l_W)+[(1+{2\over
t_W})
tan^2\theta_W-(5+{2\over t_W})]I_1(t_W,l_W)] \ , \eq\\
where $t_t =4m^2_t/\mh2$, $t_W=4\mwd/\mh2$ as before, and
$l_t=4m^2_t/ \mzd$ , $l_W=4\mwd/ \mzd$. In (20,21) the
definitions \\
\bq   I_1(a,b)={ab\over 2(a-b)}+{a^2b^2\over
2(a-b)^2}[f(a)-f(b)]+{a^2b\over(a-b)^2}[g(a)-g(b)] \  ,   \eq\\
\bq   I_2(a,b)=-{ab\over2(a-b)}[f(a)-f(b)]\ \ \ \ , \  \eq\\
where $f(t)$ is given in (7) and\\
\bqa
\ \ \ \ \ g(t)=\sqrt{t-1}sin^{-1}({1\over\sqrt{t}}) &
\makebox{ \ \ \ if \ \ \ }  &
t\geq 1 \ \ \ , \ \nonumber \\[.2cm]
g(t) ={1\over2}\sqrt{t-1}\left[ln\left({1+\sqrt{1-t}
\over1-\sqrt{1-t}}\right)-i\pi\right] & \makebox{\ \ \ if \ \ \ }
& t< 1 \ \ \ \ \ . \ \ \
\eqa \par

Finally the TT and LL components of the $H \to ZZ$ width
are given by\\
\bq \Gamma(Z_TZ_T)= \left({\alpha \beta_Z \mwd \over 4 m_H
 \swd c^4_W}\right) \left[1-d{c^4_W \over
M^2_W}(m^2_H-2M^2_Z) \right]^2 \ \ \ \ , \ \ \ \ \     \eq\\
\bq \Gamma(Z_LZ_L)= \left({\alpha \beta_Z \over 8 \swd
 m_H}\right) \left[{(\mh2-2M^2_Z)^2 \over 4 \mwd}+4d^2\mwd-
 2d(\mh2-2M^2_Z)\right]\ \ \ \ \ . \ \ \ \    \eq\\
with $\beta_Z=\sqrt{1-4M^2_Z/\mh2}$.
As in the WW case it is mainly the TT part which is affected
by $d$, while the total $H \to ZZ$ width given by\\
\bqa \Gamma_{tot}(ZZ) = \ \ \ \ \ \ \ \ \ \ \ \ \ \ \ \ \ \ \ \
\ \ \ \ \ \ \ \ \ \ \ \ \ \ \ \ \ \ \ \ \ \ \ \ \ \ \ \ \ \ \ \
\ \ \ \ \ \ \ \ \ \ \  \ \
\ \ \ \ \ \ \ \ \ \ \ \ \ \ \ \ \ \nonumber\\[0.2cm]
  \left( {\alpha  \beta_Z \over 8 \swd m_H}\right )
\left [\frac{2\mwd}{c^4_W}+ {(\mh2-2\mzd)^2\over
4\mwd}
 + 4d^2\left \{c^4_W{(\mh2-2\mzd)^2\over 2\mwd}+\mwd\right\}-
6d(\mh2-2\mzd) \right ]
\ ,  \eqa\\
is less sensitive.\par

Results are presented in Fig.2a-c and 3a-c. In the standard case
the cross section of
$e^+e^- \to \nu \bar\nu (WW) \to \nu \bar\nu H$ is about 10 times
larger than the one of
$e^+e^- \to e^+e^- (V_1V_2) \to e^+e^- H$. However as soon as $|d|$
reaches values of the order of 0.01, because of the large contributions
from the $H\gamma \gamma$ and $H\gamma Z$ couplings, both
processes $e^+e^- \to \nu \bar\nu H$ and
$e^+e^- \to e^+e^- H$
become of
the same order of magnitude. The level of
sensitivity to $d$ that can be roughly expected is $\sim 0.01$.
It is weaker than the one from laser $\gamma\gamma$ collisions
because of the suppressed
Weiszecker-Williams luminosities \cite{fv},
and the fact that WW and ZZ fusion already occurs at tree
level in SM.\par

\section{Associate production in $e^+e^- \to HZ$ and $e^+e^- \to H
\gamma$}

 The process $e^+e^- \to HZ$ is the main one \cite{LEP2HZ}
planed to be used for Higgs search in
$e^+e^-$ collisions beyond the Z peak at LEP2 or a 0.5 TeV
Collider.
The cross
section for this process decreases with energy, but the
rate is still acceptable at NLC;
see \cite{abdel}.
The relevant interactions are given by the $HZZ$ coupling
obtained by adding the tree  SM and $\O_{UW}$ contributions,
and the $H\gamma Z$ coupling derived by adding the 1-loop SM
and tree level $\O_{UW}$ results. In order to discuss the $d$
sensitivity it is sufficient to have the tree level result\\
\bqa   \sigma(e^-e^+ \to HZ) = {\pi \alpha^2 \mwd \beta_H
\over \swd}
{}~\{(1+{s \beta^2_H\over12\mzd})~
{a^2_Z+b^2_Z\over c^4_W(s-\mzd)^2} \nonumber \\[0.5cm]
  +2d\,{(s+\mzd-\mh2)\over \mwd}~[{a^2_Z+b^2_Z\over(s-\mzd)^2}
-{a_Zs_W\over c_Ws(s-\mzd)}] \nonumber \\[0.5cm]
   +4d^2\,[({s+\mzd-\mh2\over2\mwd})^2-{s^2 \beta^2_H\over12m^4_W}]
[{(a^2_Z+b^2_Z)c^4_W\over(s-\mzd)^2}-{2a_Zc^3_Ws_W\over s(s-\mzd)}
+{c^2_Ws^2_W\over s^2}] \}\ ,  \eqa\\
where $a_Z=(-1+4\swd)/(4s_Wc_W)$ ,
$b_Z=-1/(4s_Wc_W)$ and $\beta_H=2p_H/\sqrt{s}$ with $p_H$
denoting the momentum of the final Higgs.\par

The sensitivity to $d$ comes from the linear term associated to the
$HZZ$
coupling and from the quadratic terms associated to both $HZZ$ and
$H\gamma Z$ couplings. The result is shown in Fig.4a-c.
Because of the rather low rate at high energy, one cannot expect this
process to compete with the laser $\gamma \gamma$, especially
for high Higgs masses. Nevertheless a $d$ sensitivity
of $\pm 0.005$ may be reached this way
for low  $m_H$, in $e^+e^-$ collisions at NLC 0.5TeV.
The corresponding result at LEP2 (190 GEV)
with an integrated luminosity of $500pb^{-1}$
is $\pm 0.01$ for $M_H = 80 GeV$.\par

We have also looked at the process $e^+e^- \to H\gamma$, which receives
no tree SM contribution. At 1-loop, this process goes
through photon exchange due to the $H\gamma\gamma$ coupling and
Z exchange due to the $HZ\gamma$ coupling. In addition gauge
boson box diagrams also contribute \cite{hgg}. The
resulting SM cross section beyond the Z peak is too low to be
observable. However when the tree level $H\gamma\gamma$ and
$HZ\gamma$ couplings due to $\O_{UW}$ are included
it becomes significant. This later tree level result is\\
\bq   \sigma(e^-e^+\to H \gamma) =
{2\pi\alpha^2 s^2 \over 3\swd \mwd}\beta^3_H d^2
[{(a^2_Z+b^2_Z)c^2_W s^2_W\over (s-\mzd)^2}
 -{2a_Z c_W s^3_W\over s(s-\mzd)} +{s^4_W \over s^2}]
\ ,  \eq\\
 on the basis of which only a few events
would be expected  at LEP2 if
$|d|>0.05$. At a higher energy this $|d|$ limit
may reach 0.01.\par

Similar types of processes are available at pp colliders.
For example the $q\bar q' \to WH$ cross section is given by\\
\bqa   \sigma(q\bar q' \to HW) =
{N_c\pi \alpha^2 m^2_W\over 4s^4_W(s-M^2_W)^2}\beta_H
[(1+{s \beta^2_H\over
12m^2_W}) \nonumber \\
  +2d{s+m^2_W-m^2_H\over m^2_W}
   +4d^2[({s+m^2_W-m^2_H\over2m^2_W})^2-{s^2 \beta^2_H\over12m^4_W}]]
\ .  \eqa\\
Here the bare sensitivity to $d$ is similar to the one
in $e^+e^- \to HZ$, however
the experimental accuracy and the presence of large
backgrounds can certainly not allow to reach the values
obtained above.

\section{Tests with ratios of Higgs partial decay widths}

We now turn to another way of testing $d$.  Once the H is
discovered and its mass known, using any copious way of producing it
one can study how the ratios of various decay modes are sensitive to
$d$.
This is exactly how one should check that a newly discovered
scalar particle is a candidate for a Higgs boson.
 One has to verify that all its couplings agree with the
standard prediction that they should be
proportional to the mass of the particle
to which it couples.\par

We have already given in (4,19,27,18) the H decay widths to
$\gamma \gamma$, $\gamma Z$, $ZZ$ and $W^+W^-$ respectively, including
SM and $\O_{UW}$ contributions. If the Higgs mass is smaller
than twice the gauge boson mass (but larger than one boson mass)
we need to compute also the decay into one real and one virtual
gauge boson. The results for SM have been given in
\cite{Marciano}. Including also $\O_{UW}$ contributions
we get \\
\bq
\Gamma (H \to W^*W)~=~\frac{3\alpha^2m_H}{32 \pi s^4_W}~
[D_{SM}(x)+dD_1(x)+8d^2D_2(x)]
\ \ \ \ \ ,\ \ \ \ \  \eq\\
 \bq
\Gamma (H \to Z^*Z)=\frac{\alpha^2m_H}{128 \pi s^4_W c^4_W}~
\left(7-\frac{40\swd}{3}+\frac{160s^4_W}{9}\right)
[D_{SM}(x)+dc^2_WD_1(x)+8d^2c^4_WD_2(x)]
\ \ ,  \eq\\
where\\
\bqa
D_{SM}(x)~=~\frac{3(20x^2-8x+1)}{\sqrt{4x-1}}cos^{-1}
\left(\frac{3x-1}{2x^{3/2}}\right) \nonumber \\[.5cm]
\null  -(1-x)\left(\frac{47x}{2} -
\frac{13}{2} +\frac{1}{x} \right) -
3(2x^2-3x+\frac{1}{2})lnx \ \ \ \ \
\ , \ \eqa\\
\bqa
D_1(x)~=~\frac{24(14x^2-8x+1)}{\sqrt{4x-1}}cos^{-1}
\left(\frac{3x-1}{2x^{3/2}}\right) \nonumber \\[.5cm]
+12(x-1)(9x-5)-12(2x^2-6x+1)lnx \ \ \ \ \ , \ \ \
\eqa\\
\bqa
D_2(x)~=~\frac{54x^3-40x^2+11x-1}{x\sqrt{4x-1}}cos^{-1}
\left(\frac{3x-1}{2x^{3/2}}\right) \nonumber \\[.5cm]
+\frac{(x-1)}{6}(89x-82+\frac{17}{x})-(3x^2-15x+\frac{9}{2}-
\frac{1}{2x})lnx
\ \ \ \ \ , \eqa\\
and $x=(M_V/m_H)^2$ with $M_V=M_W$ or $M_Z$.
These results are particularly important
for $90GeV\leq m_H \leq 140 GeV$.\par

Finally we also need for comparison the fermionic Higgs
decay widths, which are not affected by $\O_{UW}$ at the tree
level. Let us take $H \to
b \bar b$ as a reference, since the $t \bar t$ threshold
is probably very
high. The $H \to b \bar b$ partial width, which is purely
standard and particularly
important if $m_H\lsim 140GeV$, is\\
 \bq
\Gamma(H\to b\bar b)= 3{\sqrt2G_F m^2_b\over8\pi}\beta^3_b m_H \ \
 \ \ \ . \ \ \
  \eq\par
with $\beta_b = \sqrt{1-{4m^2_b/m^2_H}}$.

In Fig.5a-e we show a sample of ratios of Higgs decay
widths, as functions of the Higgs mass and the $d$ coupling.
Independently of the Higgs mass, a large
sensitivity to $d$ exists for
the $H \to \gamma \gamma$ decay width, as opposed to $H
\to WW (WW^*)$, $H \to ZZ (ZZ^*)$ or $H \to b \bar b$
in Fig.5a-d. A sharp and
fortuitous SM-$d\O_{UW}$ interference occurs for $d=0.01$ and
$m_H=0.15TeV$.
$H \to \gamma Z$  is also very sensitive to d. On the
opposite the WW and ZZ channels are less sensitive as one could guess
from (18,27,31,32) and this sensitivity becomes even smaller in
the ratio WW over ZZ.\par

  The use of these results for giving limits on d will actually depend
on the number of events that will be observed. This test can
however be performed independently of the production mode and one can
cumulate events from various Higgs boson sources.\par

\section{Unitarity constraints on $d$ from various Higgs production
channels}

In this Section we give the explicit check that the unitarity
constraints on $d$ established in \cite{uni}
using $J=0$ partial
waves, are still valid when considering the $J=1$ case also.
Generally, because of the $2J+1$ weight factor, higher partial waves
give weaker constraints than lower ones. However the point
is that for $J=1$, new channels containing the Higgs are
involved that did not contribute to $J=0$ amplitudes.\par

Following the same procedure as in \cite{uni},
we use the high energy
expressions of the scattering amplitudes given in \cite{VVVV}
as well as the
ones involving one or two Higgs states given in Appendix A and B.
We have then to separately treat three sets
of VV, VH, HH coupled channels, with
total charge $Q=2,1,0$.\par

 The $Q=2$, $J=1$ set involves five independent non vanishing
$W^+W^+$ scattering amplitudes. It only contains terms linear
in $d$. Diagonalizing the relevant $5\times 5$ matrix and demanding
that the largest eigenvalue is less than 2 gives the unitarity
constraint\\
\bq
|d|\lsim {24s^2_W M^2_W \over s\alpha} \simeq
780{M^2_W\over s}\ \ \ \ \ \ . \ \ \ \
\eq\par

The $Q=1$, $J=1$ set involves 13 different $ZW$, $\gamma W$ and $HW$
states. The corresponding $13\times 13$ matrix is rather
involved and contains both $d$ and $d^2$
terms. It produces a complicated analytical constraint which
is numerically approximated by \\
\bq
|d| \lsim 25.4{M^2_W\over s} + 9.68{M_W\over\sqrt{s}}
\ \ \ \ \ \ \ . \ \ \ \ \  \eq \par

Finally, the $Q=0$, $J=1$ set a priori involves
26 different $W^+W^-$,
$ZZ$, $\gamma
\gamma$, $\gamma Z$, $H \gamma$, $HZ$ and $HH$ states. However only 12
of them contribute to $J=1$ namely,
7 $W^+W^-$, 3 $HZ$ and 2 $H\gamma$. The relevant
$12\times 12$ matrix leads again to a complicated unitarity
constraint which is numerically approximated by\\
\bq
|d| \lsim 0.67{M^2_W\over s} + 27.04{M_W\over\sqrt{s}}
\ \ \ \ \ \ . \ \ \ \ \ \ \ \ \ \eq\\
One can check that in the TeV range the constraints
(34-36) turn out to be
less stringent
than the one established from the $J=0$ partial waves in \cite{uni}.
Only for energies larger than 30 TeV, the constraint (37)
coming from the
$Q=2$, $J=1$ set, which is linear in $d$,
becomes more stringent.\par

\section{Final discussion}

 In this paper we have shown that limits on the H and W
interactions due to the $SU(2)_c$ symmetric operator  $d\O_{UW}$,
can be largely improved by considering direct H production
processes. The best process seems to be the Higgs production in
photon-photon collisions from laser backscattering at a high energy
$e^+e^-$ linear collider. With the expected
luminosities, an observability
limit for $d$ of the order of $\pm0.001$ can be achieved
 for a large range
of Higgs masses. The other
boson-boson fusion processes are less efficient because of the
suppressed Weiszecker-Williams luminosities  and they would only be
sensitive to $d$ values at the level of 0.01.
For low and intermediate Higgs masses,
the associate production $e^+e^- \to HZ$ is an interesting
possibility which could lead to a 0.005 limit. Even at LEP2 a 0.01
limit for d should be settled , provided $m_H \sim 80 GeV$.
The other process
 $e^+e^- \to H\gamma$ is too weak to be competitive and can hardly
reach 0.01 at NLC and 0.05 at LEP2.\par

  We have also shown how ratios of Higgs partial decay widths
could reflect the same
sensitivity to d, provided one could accumulate a sufficient number
of Higgs events.    The resulting
panorama of the expected sensitivities for $d$ from all
direct and indirect processes involving Higgs and gauge bosons
is given in Table 1.\\

 \begin{center}
\begin{tabular}{|c|c|c|} \hline
\multicolumn{3
}{|c|}{Table 1: Sensitivity to $d\O_{UW}$}
\\ \hline
\multicolumn{1}{|c|}{process} &
  \multicolumn{1}{|c|}{collider} &
   \multicolumn{1}{|c|}{ $|d|$} \\ \hline
 pp($q \bar q$, WW fusion) & LHC & 0.1 \\
 $\gamma \gamma \to WW,ZZ$ & NLC 0.5-2 TeV & 0.25-0.02 \\
 $\gamma \gamma \to H$ & NLC 0.5-2 TeV &  0.001 \\
 $WW,ZZ \to H$ & NLC 0.5-2 TeV &  0.01 \\
 $e^+e^- \to HZ$ & NLC 0.5 TeV &  0.005 \\ \hline
\end{tabular}
\end{center}

\noindent
We observe that a gain of a factor 10 to 100 is obtained for the
$d$ sensitivity from
direct Higgs boson production, as compared to the
sensitivity expected  from
processes involving only gauge bosons in the final state.
The reason for this is beacuse the later processes depend on
$d$ only through Higgs exchange diagrams \cite{VVVV},
\cite{ZepHiggs}.\par

  It is also interesting to compare these results to theoretical
expectations. In this paper we have confirmed the validity
of the unitarity
limits announced in a previous letter. Using the full set of coupled
VV, VH and HH channels we have shown that the strongest constraints
on $d$, for an NP scale lying in the TeV range, indeed
come from the $J=0$ partial waves and can be written as\\
\bq
 |d|\ \lsim \  17.6\ \frac{M_W}{s}\ + \ 2.43 \ \frac{M_W}{\sqrt s}
 \
\ \ \ \ . \ \ \ \eq\par

It appears therefore that the level of the achievable
sensitivity from the
processes studied in this paper is largely within the
 domain allowed by unitarity.
Such tests will therefore produce essential informations on the
possible New Physics effects affecting the scalar sector
and described by the
$SU(2)_c$ symmetric operator $\O_{UW}$.
The 0.01 to 0.001 level of
observability expected
for the coupling $d$, is
comparable to the one expected for the coupling $\lambda_W$
describing the $SU(2)_c$ symmetric interactions among W
bosons only \cite{nlcgr}.
This level is the one at which standard electroweak radiative
corrections start to contribute. It makes therefore sense to
persue high precision tests of the presence of such
interactions.
The resulting constraints that will be put on the gauge sector and on
the scalar sector of the SM will turn out to be as stringent as those
which have been obtained on the fermionic sector at LEP1. The
combination of all these information should be essential in order
to select the allowed ways to extend the SM
and to cure its deficiencies.\par

{\Large \bf Acknowledgements}\par
 It is a pleasure to thank A. Djouadi for many useful
discussions and informations
about Higgs boson phenomenology.
\newpage

\def\cw{c_W}
\def\sw{s_W}
\def\swd{s^2_W}
\def\cwd{c^2_W}
\def\cot{\cos\theta}
\def\cott{\cos^2\theta}
\def\sit{\sin\theta}
\def\ed{e^2}
\def\mwd{M^2_{W}}
\def\mw{M_W}
\def\mwt{M^3_{W}}
\def\lw{\lambda_W}
\def\rd{\sqrt{2}}
\def\rs{\sqrt{s}}

\renewcommand{\theequation}{A.\arabic{equation}}
\setcounter{equation}{0}
\setcounter{section}{0}

{\Large\bf Appendix A : Helicity amplitudes for \underline{single}
\underline{Higgs} production in boson fusion
processes at high energy}\\

The  $V_1(\lambda)V_2(\tau) \to H V_3(\mu)$ processes are described
by 27 helicity
amplitudes $F_{\lambda \tau \mu}(\theta)$, where $\lambda$, $\tau$,
and $\mu$ denote the helicties,
$\theta$ is the c.m.\@ angle between $V_1$ and H, and
the normalization is
such that the differential cross section writes as\\
\bq
   {d\sigma(\lambda \tau \mu)\over d cos(\theta)} =
   C|F_{\lambda \tau \mu}(\theta)|^2 \ \ \ \ \ , \ \ \ \ \  \
  \eq\\
where\\
\bq  C = {1\over32\pi s}\,{p_{H}\over p_{12}} \ \ \ \ \ \ \
\ \ \eq\\
includes \underline{no spin average}.\par

\section{Standard, $\O_W $ and $\O_{UW}$ contributions}

In the following expressions the last factor of the type (a,b)
always refers to (Z,$\gamma$) production respectively.\par
\vspace{0.5cm}
\fbox{ {\bf $W^-W^+ \to HZ, H \gamma $ }}\par
\bq
F_{\pm  \mp,\mp }=\pm\biggm\{{\ed d \rs \over \rd \mw} \sit {(1+\cot)
 \over (1-\cot)} - {\ed \lw d s^{3/2} \over 2 \rd \mwt}
\sit \biggm\} \left({\cw  \over \swd} , {1 \over \sw}\right)
\eq\\
\bq
F_{\pm \mp, \pm}=\pm\biggm\{{\ed d \rs \over \rd \mw} \sit{(\cot -1)
\over (1+\cot)} - {\ed \lw d s^{3/2} \over
2\rd\mwt}\sit\biggm\} \left({\cw
 \over \swd}, {1\over \sw} \right )\eq\\
\bq
F_{\pm \pm,\pm}=\pm\biggm\{{4\ed d \rs \over \rd\mw\cw}{1 \over
\sit} + {\ed \lw d s^{3/2}\sit \over 2\rd\mwt}\biggm\}
\left({\cw \over \swd}, {1 \over \sw}\right)\eq\\
\bq
F_{\pm \pm,\mp}=\mp\biggm\{{\ed\lw d s^{3/2} \sit \over
2\rd\mwt}\biggm\} \left({\cw \over \swd},{1\over \sw}\right)\eq\\
\bq
F_{0 \mp,\mp}=\biggm\{{\ed(1-4\cwd-\cot)\over
2\cw\swd(\cot-1)}\biggm\}\left(1, {\sw\over\cw}\right)\eq\\
\bq
F_{\pm 0,\pm}=\biggm\{{\ed(-1+4\cwd-\cot)\over
2\cw\swd(1+\cot)}\biggm\} \left(1,{\sw\over\cw}\right)\eq\\
\bq
F_{0\pm,\mp}={\ed\lw s\over
8\mwd}(3+\cot)\left({\cw\over\swd},{1\over\sw}\right)\eq\\
\bq
F_{\pm 0,\mp}={\ed\lw s\over
8\mwt}(3-\cot)\left({\cw\over\swd},{1\over\sw}\right)\eq\\
\bq
F_{\pm\mp,0}=-{\ed\over 2\swd}\cot(1,0)\eq\\
\bq
F_{\pm\pm,0}=-{\ed\lw s\over 4\swd\mwd}\cot(1,0)\eq\\
\bq
F_{00,\mp}=\pm{\ed d \rs\over 2\rd\mw}\sit\left({\cw\over
\swd},{1\over\sw}\right)\eq\\
\bq
F_{0\mp,0}=\pm{\ed d\rs\over \rd\swd\mw}{\sit \over (1+\cot)}(1,0)
\eq\\
\bq
F_{\pm 0,0}=\pm{\ed d\rs\over \rd\swd\mw}{\sit \over (1-\cot)}(1,0)
\eq\\
\bq
F_{00,0}={\ed(1-10\cwd-\cott(1-2\cwd)\over
4\cwd\swd(\cott-1)}\cot(1,0)\eq\\
\null\vspace{1.0cm}\par

\fbox{ {\bf $W^-Z,W^-\gamma \to HW^- $ }}\par
\bq
F_{\pm\pm,\pm}=\mp\biggm\{{4\ed d\rs\over \rd\mw}{1\over \sit}-
{\ed\lw d s^{3/2} \over
2\rd\mwt}\biggm\}\left({\cw\over\swd},{1\over\sw}\right)\eq\\
\bq
F_{\pm\mp,\pm}=\pm\biggm\{{\ed d\rs \over \rd\mw}\sit{(1-\cot)\over
(1+\cot)}+{\ed\lw d s^{3/2} \over
2\rd\mwt}\biggm\}\left({\cw\over\swd},{1\over\sw}\right)\eq\\
\bq
F_{\pm\mp,\mp}=\mp\biggm\{{\ed d\rs\over \rd\mw}\sit{(1+\cot)\over
(1-\cot)}+{\ed\lw d s^{3/2} \over
2\rd\mwt}\biggm\}\left({\cw\over\swd},{1\over\sw}\right)\eq\\
\bq
F_{\pm\pm,\mp}=\pm{\ed\lw d s^{3/2} \over
2\rd\mwt}\sit\left({\cw\over\swd},{1\over\sw}\right)\eq\\
\bq
F_{0\pm,\pm}={\ed(1-4\cwd)(1+\cot)+\cot-3\over
4\cw\swd(1-\cot)}\left(1,{\sw\over\cw}\right)\eq\\
\bq
F_{\pm\mp,0}=-{\ed(1-\cwd(1+\cot))\over
2\cw\swd}\left(1,{\sw\over\cw}\right)\eq\\
\bq
F_{0\pm,\mp}={\ed\lw s\over
8\mwd}(3+\cot)\left({\cw\over\swd},{1\over\sw}\right)\eq\\
\bq
F_{\pm\pm,0}=-{\ed\lw s\over
4\mwd}\cot\left({\cw\over\swd},{1\over\sw}\right)\eq\\
\bq
F_{\pm0,\pm}=-{\ed\over 2\swd}{(3-\cot)\over (1+\cot)}(1,0)\eq\\
\bq
F_{\pm0,\mp}={\ed\lw s\cw\over 8\swd\mwd}(3-\cot)(1,0)\eq\\
\bq
F_{00,\pm}=\pm{\ed d\rs\over 2\rd\swd\mw}\sit(1,0)\eq\\
\bq
F_{0\pm,0}=\pm{\ed d\rs\cw\over\rd\swd\mw}{\sit \over (1+\cot)}
(1,0)\eq\\
\bq
F_{\pm0,0}=\pm{\ed d\rs\over\rd\swd\mw}{\sit \over(\cot-1)}(1,0)\eq\\
\bq
F_{00,0}={\ed(3-\cot)[\cot(1-4\cwd)-\swd-\cwd\cott]\over 4\cwd\swd
(\cott-1)}(1,0)\eq\\
%\newpage

\renewcommand{\theequation}{B.\arabic{equation}}
\setcounter{equation}{0}
\setcounter{section}{0}

{\Large\bf Appendix B : Helicity amplitudes for \underline{double}
\underline{Higgs} production in boson fusion processes at high energy
}\\

    These  $V_1(\lambda)V_2(\tau) \to HH$ processes are described by
9 helicity amplitudes  $F_{\lambda \tau}(\theta)$.
$\theta$ is the angle between $V_1$ and H and the normalization is
such that the differential cross section writes\\
\bq     {d\sigma(\lambda \tau )\over d cos(\theta)} =
 C|F_{\lambda \tau}(\theta)|^2  \ \ \ \ \ , \ \ \ \eq\\
where\\
\bq  C = {1\over32\pi s}\,{p_{H}\over p_{12}} \ \ \ \ \ , \ \ \
 \eq\\
includes \underline{no spin average}.\par
\vspace{1.0cm}

\fbox{ {\bf $W^-W^+, ZZ, \gamma\gamma, \gamma Z \to HH $ }}\par
\bqa
 F_{\lambda \tau}(\theta) = -(1-\delta_{\tau 0})
(1-\delta_{\lambda 0})(1, \cwd, \swd,
\sw \cw) . \nonumber \\[0.5cm]
\biggm\{{d^2g^2_2 s\over 2\mwd}(1+3\lambda\tau)+{dg^2_2s\over
4\mwd}(1+\lambda\tau)\biggm\}
\eqa\\
\null\vspace{1cm}

\fbox{ {\bf $WH \to HW,\, ZH \to HZ,\, \gamma H \to H\gamma,
\, \gamma H \to HZ $ }}\par
\vspace{0.5cm}
These $V_1(\lambda)H \to HV_2(\mu)$ channels are obtained by
crossing those above. The helicity amplitudes are now
given by\\
\bqa
F_{\lambda \mu}(\theta) = -(1-\delta_{\mu 0})
(1-\delta_{\lambda 0})(1, \cwd, \swd,
\sw \cw)(1+\cot) . \nonumber \\[0.5cm]
\biggm\{{d^2g^2_2 s\over
4\mwd}(1-3\lambda\mu)+{dg^2_2s\over
8\mwd}(1+\lambda\mu)\biggm\}
\eqa

\newpage

\centerline { {\bf Figure Captions }}\par
 Fig.1 Cross sections for Higgs production in
$\gamma\gamma$ collisions
from laser backscattering at a 0.5 TeV (a), 1 TeV (b), 2 TeV (c)
  $e^+e^-$ linear collider. Standard prediction
(solid line),
 with $d =+ 0.01$ (long dashed),  $d =- 0.01$ (dashed - circles),
  $d =+ 0.005$ (short dashed),  $d =- 0.005$ (dashed),
  $d =+ 0.001$ (dashed - stars), and $d =- 0.001$ (dashed - boxes).
 The expected number of events per year
for an integrated luminosity of $10 fb^{-1}$ is also
indicated. \\
\null\\
 Fig.2 Cross sections for Higgs production in $e^+e^- \to H \nu\bar
\nu$
through $WW$ fusion  at a 0.5 TeV (a), 1 TeV (b), 2 TeV (c)
 $e^+e^-$ linear collider. Standard prediction
(solid line),   with $d =+ 0.01$ (short dashed),
 $d =- 0.01$ (dashed - circles),
 $d =+ 0.05$ (long dashed),  $d =- 0.05$ (dashed - circles),
 $d =+ 0.1$ (dashed - stars),  and $d =- 0.1$ (dashed - boxes).\\
\null\\
 Fig.3 Cross sections for Higgs production in $e^+e^- \to H e^+e^-$
through $\gamma \gamma$, $\gamma Z$ and $ZZ$ fusion
at a 0.5 TeV (a), 1 TeV
(b), 2 TeV (c)
 $e^+e^-$ linear collider. Standard prediction
(solid line),  with $d =+ 0.01$ (short dashed),
 $d =- 0.01$ (dashed - circles),
 $d =+ 0.05$ (long dashed), $d =- 0.05$ (dashed - circles),
 $d =+ 0.1$ (dashed - stars), and $d =- 0.1$ (dashed - boxes).\\
\null\\
 Fig.4 Cross sections for associate Higgs production
 in $e^+e^- \to HZ$  at a 0.5 TeV (a), 1 TeV (b), 2 TeV (c)
 $e^+e^-$ linear collider. Standard prediction
(solid line),  with $d =+ 0.01$ (long dashed),  $d =- 0.01$
(dashed - circles),
  $d =+ 0.005$ (short dashed), with $d =- 0.005$ (dashed),
 $d =+ 0.001$ (dashed - stars), and $d =- 0.001$ (dashed -
boxes).\\
\null\\
 Fig.5 Ratios of Higgs decay widths
$\Gamma(H\to\gamma \gamma)/\Gamma(H\to b  \bar b)$ (a),
$\Gamma(H\to\gamma \gamma)/\Gamma(H\to WW)$ (b),
$\Gamma(H\to\gamma \gamma)/\Gamma(H\to ZZ)$ (c),
$\Gamma(H\to\gamma \gamma)/\Gamma(H\to \gamma Z)$ (d).
 Standard prediction (solid line),
with $d =+ 0.01$ (long dashed), $d =- 0.01$
(dashed - circles),
 $d =+ 0.005$ (short dashed), $d =- 0.005$ (dashed),
 $d =+ 0.001$ (dashed - stars), $d =- 0.001$
(dashed - boxes).


\newpage

\begin{thebibliography}{99}
%
\bibitem{nlc} Proc.Workshop on Physics at Future Accelerators,
 La Thuile,
J.H.Mulvey, ed. CERN Report 87-07;
 Report, Opportunities and Requirements for Experimentation at a Very
High Energy $e^+e^-$ Collider, SLAC-329(1928);
 Proc. Workshops on Japan Linear Collider, KEK Reports, 90-2, 91-10
and 92-16;
 Physics and experiments with $e^+e^-$ Linear Colliders, Saariselk\"a
1991, R.Orava, P.Eerola and M.Nordberg eds., World Scientific,
Singapore 1992.
%
\bibitem{lhc}Proc. of the Large Hadron Collider
Workshop,ed.G.Jarlrskog and D.Rein, CERN 90-10,
ECFA 90-133(1990).
%
\bibitem{phys} P.M.Zerwas, DESY 93-112, Aug.1993;
  Proc. of the Workshop on
    $e^+e^-$ Collisions a  500 GeV: The Physics Potential,
    DESY 92-123A+B(1992), ed. P.Zerwas;
%
\bibitem{VVVV} G.J.Gounaris and F.M.Renard, \zp{C59}{1993}{143};
erratum, p.682.
%
\bibitem{uni} G.J.Gounaris, J.Layssac and F.M.Renard,
preprint PM/93-37, THES-TP 93/11, revised version January 1994.
%
\bibitem{LHCGLR} G.J.Gounaris, J.Layssac and F.M.Renard, preprint
   PM/93-26, THES-TP 93/8, to be published in Zeit. Phys.C.
%
\bibitem{nlcgr}  G.J.Gounaris and F.M.Renard, preprint PM/94-03,
THES-TP 94/01, to appear in Phys.Lett.B.
%
\bibitem{SU2c} G.J.Gounaris and F.M.Renard, \zp{C59}{1993}{133}.
%
\bibitem{Bil} M.Bilenky, J.L.Kneur, F.M. Renard and D.
Schilknecht,\np{B409}{1993}{22}.
%
\bibitem{ZepHiggs} A.De Rujula et al, \np{B384}{1992}{3}.
K.Hagiwara et al, \pl{B283}{1992}{353};\pr{D48}{1993}{2182}.
K.Hagiwara et al, \pl{B318}{1993}{155}.
%
\bibitem{hgg} L.Bergstr\"om and G.Hulth, \np{B276}{1986}{744}
%
\bibitem{laser}  I.Ginzburg et al, Nucl. Instrum. Methods
${\bf 205}$(1983)47; ${\bf 219}$(1984)5.
%
\bibitem{fgamma} V.Telnov, Nucl. Instrum. Methods ${\bf A294}$(1990)72.
%
\bibitem{Kuhn} J.H.K\"{u}hn, E.Mirkes and J.Steegborn,
\zp{C57}{1993}{615}.
%
\bibitem{abdel} A. Djouadi, preprint Montreal University 1994.
%
\bibitem{fv} M.Capdequi Peyranere et al, \zp{C41}{1988}{99}.
%
\bibitem{LEP2HZ} The Higgs Hunter's Guide, J.F.Gunion et al, Redwood
City (1990).
%
\bibitem{Marciano} W-Y Keung and W.J. Marciano, \pr{D30}{1984}{248}.


\end{thebibliography}
\end{document}